\documentclass[manuscript]{aastex}

\usepackage{color}

\shorttitle{Turbulence decay and cloud core relaxation in molecular clouds}
\shortauthors{Y. Gao, H. Xu, C. K. Law}

\begin{document}


\title{Turbulence decay and cloud core relaxation in molecular clouds}


\author{Yang Gao\altaffilmark{1, 2, 3}}
\author{Haitao Xu\altaffilmark{4}}
\author{Chung K. Law\altaffilmark{1, 3}}
\affil{$^1$ Center for Combustion Energy, Tsinghua University, Beijing 100084, China}
\affil{$^2$ Department of Thermal Engineering, Tsinghua University, Beijing 100084, China}
\affil{$^3$ Department of Mechanical and Aerospace Engineering, Princeton University,\\ Princeton, New Jersey 08544, USA}
\affil{$^4$ Max Planck Institute for Dynamics and Self-Organization (MPIDS),\\ 37077 G\"{o}ttingen, Germany}







\begin{abstract}
The turbulent motion within molecular clouds is a key factor controlling star formation.
Turbulence supports molecular cloud cores from evolving to gravitational collapse
  and hence sets a lower bound on the size of molecular cloud cores in which star formation can occur.
On the other hand, without a continuous external energy source maintaining the turbulence,
  such as in molecular clouds, the turbulence
  decays with an energy dissipation time comparable to the dynamic timescale of clouds,
  which could change the size limits obtained from Jean's criterion by assuming constant turbulence intensities.
Here we adopt scaling relations of physical variables in decaying turbulence to
  analyze its specific effects on the formation of stars.
We find that the decay of turbulence provides an additional approach for Jeans' criterion to be achieved,
  after which gravitational infall governs the motion of the cloud core.
This epoch of turbulence decay is defined as cloud core \emph{relaxation}.
The existence of cloud core relaxation provides a more complete understanding in the competition between turbulence and gravity on
  the dynamics of molecular cloud cores and star formation.

\end{abstract}





\section{Introduction}

Turbulent motion, magnetic field and gravity are governing factors on the dynamics of star formation
 in molecular clouds \citep{pudritz2002,wardthompson2002,mckee2007}.
Turbulence and its effects on molecular clouds and star formation have been studied since the pioneer work of
  \cite{chandrasekhar1949,chandrasekhar1951a,chandrasekhar1951b}; a recent review on this topic is given in \cite{maclow2004}.
Broadly speaking, turbulence has two competing effects on star formation. On one hand, large-scale turbulence is the main driving mechanism
  that creates dense cloud cores~\citep{chandrasekhar1951a,larson1981,kritsuk2013}, which incubate star formation.
On the other hand, turbulent motion within the cloud cores provides additional support against gravitational collapse
  ~\citep{chandrasekhar1951b,bonazzola1987,leorat1990}, which hinders star formation.
A full understanding of the dynamics of start formation therefore requires a complete analysis of the effects of turbulence.

In this work, we examine the resistance to gravitational collapse due to the in-cloud-core turbulence.
Most of the previous analyses of this effect were based on the theory of compressible or incompressible, fully developed and
  statistically stationary turbulence, i.e., turbulent flows maintained by continuous external energy supplies \citep[][]{mckee2007}.
Most of the energy sources that drive the turbulent motion in a molecular cloud, on the other hand,
  are neither uniform nor consistent \citep[][]{maclow2004},
  which means that the turbulence in any particular cloud core cannot be continuously maintained
  and the turbulent flow will gradually slow down, i.e., the turbulent energy will decay.
Additionally, numerical simulations showed that the energy decaying time of a typical turbulent flow in molecular clouds is smaller or comparable
  to the dynamic timescale of star formation \citep{stone1998, maclow1998}.
These two factors above suggest that star formation in cloud-cores actually occurs in an environment of decaying turbulence.
Therefore, the effect of turbulent energy decay should be taken into account when analyzing star formation in cloud-cores,
  which is what we address here.

This paper is organized as follows:
  turbulence driving mechanisms in molecular clouds are first reviewed,
  followed by presentation of the scaling law of decaying turbulence,
  with an analysis of its effect on Jeans' criterion.
The epoch of cloud-core relaxation is then proposed,
  and results of our scaling analysis are discussed and concluded.

\section{Driving mechanisms of turbulence in molecular clouds}


Currently, a general consensus on star formation is that large-scale ($\sim 10-100$ pc) turbulence
  leads to the clustering of dense regions and subsequently the formation of stars.
The energy dissipation time of the large-scale turbulence is of the order of $\sim 10^6$ yr,
  which is comparable to the free fall timescale of a cloud \citep{stone1998,ossenkopf2002,offner2008}.

Various energy sources that could trigger turbulent flows in molecular clouds have been proposed and then corresponding driving scales of turbulence
  have been numerically studied \citep{genzel1998,maclow2004,joung2006,brunt2009}.
The magneto-rotational instability (MRI) is an efficient mechanism that couples large-scale galactic rotation with  turbulent motions in star forming clouds,
  whose energy input into turbulent motion is, however, about two orders of magnitude less than the observed value.
Similarly, turbulent motion due to gravitational instability (GI) is not energetic enough to drive star formation.
These two instabilities (MRI and GI) therefore likely serve as basic driving mechanisms that contribute to only
  a small portion of the observed turbulence in molecular clouds.
Protostellar jets and outflows are good sources of local turbulence drivers that can affect their surrounding cloud environment
  but are too small in dimension to account for the large-scale turbulence observed.
Massive stars can affect the cloud environment significantly by intense radiation,
  but only a small portion of the radiation energy is converted to turbulent motion.
On the other hand,
  although winds from massive stars can be more energetic,
  the population of massive star is too small to make a major contribution, especially when compared to that of supernova explosion discussed below.

It is suggested by \citet{maclow2004} that supernova explosion is the dominant turbulence driving mechanism,
  with sufficient energy input rates to trigger the turbulence observed in molecular clouds.
Following their analysis, we assume the supernova explosion rate in the Galaxy
  (100 pc star formation scale height and 15 kpc in radius)
  to be $(50~{\rm yr})^{-1}$, which then gives an estimate that the supernova explosion rate in a typical star forming cloud
  with a diameter of $100$ pc to be $\sim(10^6 - 10^7~{\rm yr})^{-1}$.
This result means that there is a typical time lag of $10^6 - 10^7$ yr between two successive supernova explosions in a molecular cloud,
  which is comparable or slightly larger than the star formation time of several $10^6$ yr.
As a result of a supernova explosion, shock waves sweep the gas and dust in the molecular cloud, cluster them into dense cores through
  particle collisions, and initiate the turbulent motion in these cores as well as in the entire cloud \citep{joung2006}.
To give a general picture of the molecular clouds and the dense cloud cores discussed here,
  the typical length scale of cloud cores is $l_{\rm core} \sim 0.1$ pc,
  which is much smaller than the cloud diameter of $l_0 \sim 100$ pc;
  and the core density is of the order $\rho_{\rm core}\sim 5\times10^4~{\rm cm}^{-3}$,
  which is much larger than the density of diffuse regions of a molecular cloud $\rho_{\rm cloud}\sim10~{\rm cm}^{-3}$
  (both are number densities of molecules).
After the shock waves pass by,
  the decay laws governs the evolution of turbulence in the molecular cloud.

\section{Turbulence decay in molecular clouds and cloud cores}


The decay of turbulent energy, and the associated variation of energy spectra in turbulent flows without an external maintaining force
  is a classical problem in fluid turbulence research. It can be dated back to the classical paper by \cite{karmanhowarth1938} and has been investigated by \cite{kolmogorov1941,batchelor1948a,batchelor1948b,heisenberg1948}, among many others.
Recent studies based on terrestrial fluid experiments improved the understanding of turbulence decay properties
 \citep{kurian2009,krogstad2010,krogstad2011};
  while in astrophysical investigations, numerical simulations of star forming clouds confirmed these scaling relations for decaying turbulence
  and expanded the results to flows with magnetic fields \citep{biskamp1999,maclow1999,cho2003}.

In this work we adopt the following decay law for incompressible turbulence, given as Eq. (A1) in \cite{krogstad2011},
  \begin{equation}
 \frac{E(t)}{E_0}=\frac{u^2(t)}{u_0^2}=\bigg(1+\frac{A}{n}\frac{u_0 t}{l_0}\bigg)^{-n},
  \label{equ:decay}
  \end{equation}
  where $E(t) = u^2(t)/2$ is the turbulent kinetic energy at time $t$ either after the start of the decaying, or the termination of external forcing, which corresponds to the passing of the shock wave in the case of energy supply by the supernova explosion.
Furthermore, $E_0$, $u_0$ and $l_0$ are the initial values, i.e., at $t = 0$, of turbulent energy, fluctuating velocity and integral scale, respectively,
  and $A$ is a dimensionless number, typically between $1/3$ and $1/2$ as found in the observation of isotropic turbulence and $A=1/2$ is used in following calculations.
Although the power-law form of decay given by Eq.~(\ref{equ:decay}) is generally accepted as universal, the exponent $n$ has not been uniquely determined.
Currently available data and theories suggest that it should be between $1$ and $2$ \citep{biskamp1999,krogstad2011},
  with many recent experimental evidences supporting that it is close to $1.2$ \citep{kurian2009,krogstad2010,sinhuber2014}.
Considering the turbulence in molecular clouds, this decay law has the following physical interpretations:
  without any external energy input maintaining the turbulence,
  the decay of the total turbulent energy follows the $\sim (1+t/t_0)^{-n}$ law
  and the slow down of the turbulent speed follows $\sim (1+t/t_0)^{-n/2}$ accordingly,
  where $t_0=l_0/u_0$ is the turbulence decay time scale.


For a (giant) molecular cloud with typical length scale $l_0 \sim 100$ pc and mass $M_0 \sim10^5~M_{\odot}$,
  if all the energy released through a typical supernova explosion $(\sim E_{\rm SN}=10^{51}~{\rm erg})$
  is converted to the turbulent energy of the cloud,
  \footnote{The cases in which 10\% and 1\% of the supernova explosion energy is converted to the turbulent energy are considered in Section 5.}
  the turbulent fluctuating velocity in the molecular cloud is $u_0=\sqrt{2E_{\rm SN}/M_0}=30~{\rm km/s}$.
The corresponding characteristic energy decay time of the turbulence in the cloud is $t_0=l_0/u_0=3\times10^6~{\rm yr}$,
  which is consistent with the simulation result of \citet{stone1998}.
Then for a typical dense cloud core of $l_{\rm core} \sim 0.1$ pc, the fluctuating velocity within the core is
  $u_{\rm core}=u_{0}(l_{\rm core}/l_0)^{1/3}=3~{\rm km/s}$ once it is formed,
  assuming the Kolmogorov scaling for incompressible turbulence in the inertial range \citep[cf.][]{kolmogorov1941}\footnote{Turbulence in molecular clouds are actually compressible thus the spectra index could be different from the Kolmogorov scaling law. Discussions on the effect of compressible turbulence can be found in the last section of this paper.}
\begin{equation}
  u\propto l^{1/3} .
  \label{equ:kolmogorov}
\end{equation}
Accordingly, the decay time scale for turbulence in such a cloud core is $t_{\rm core}=l_{\rm core}/u_{\rm core}=3\times10^4~{\rm yr}$,
  which is much smaller than the decay time of the turbulent motion in the whole cloud.
Regarding the above estimates,
  it is noted that the initial length scale of the cloud core $l_{\rm core}$ could vary for different cores,
  resulting in different core masses and turbulent speeds;
  while the initial local density of the cloud core $\rho_{\rm core}$ could also be different.

Although bounded in the gravitational potential of the large cloud,
  these dense cores form their own local potential fields and can be relatively isolated from other cores \citep{wardthompson2007}.
Due to the density difference between the dense cores and cloud diffuse regions, $\rho_{\rm core}/\rho_{\rm cloud}\sim 5\times10^3$,
  turbulent motion in the diffuse regions around a cloud core can hardly generate strong fluctuations inside the core \footnote{As an estimate, once cloud cores are formed, a turbulent speed of $u_{\rm cloud}=30$ km/s in the cloud can only result in $u_{\rm core}=0.006$ km/s flows in the dense core, which is much smaller than the inside-core turbulent motion of several km/s.}
  as a result of mass flux conservation ($\rho u=const.$).
This means that the turbulent motion within the dense cores is essentially not affected by turbulence of the diffuse cloud after the initial fluctuation,
  so the dense cores experience the decay of turbulence in a relatively isolated sense.
It has already been noticed that local star-formation behaviors are different
  for dense cores of different properties \cite[see e.g.,][]{mckee2007}:
  1) cores that have sufficiently high internal turbulent energy compared to their self-gravity potential
   will re-disperse and cannot form stars;
  and 2) cores whose turbulent energy are low enough compared to the gravitational energy will collapse under gravity and form star(s).
What we will show in the following section is that the decay of turbulent motions in the cores may allow some type 1) dense cores
  to eventually evolve to gravitational collapse.


\section{Jeans' criterion in cloud cores with decaying turbulence}

Jeans' criterion for a turbulent cloud core to be gravitationally unstable to perturbations of wave number $k$
  was derived in \cite{chandrasekhar1951b,bonazzola1987}:
  \begin{equation}
  k^2<\frac{4\pi\rho_{c} G}{c^2+\frac{1}{3}u_{c}^2},
  \label{equ:Jeans}
  \end{equation}
  where $c$ is the speed of sound, $u_c$ the turbulent speed of the cloud core
  and $G$ the gravitational constant. Hereafter variables with subscript $c$ represent properties of cloud cores.
Equation~(\ref{equ:Jeans}) shows that the gravitational instability is a long-wave instability.
In a cloud core, the core diameter $l_c$ limits the longest wave to $k=2\pi/\lambda \sim 2\pi/l_{\rm c}$,
  which is the first unstable mode when a cloud core becomes unstable due to changes in conditions.
If we take a typical cloud-core temperature of $\sim 10$ K, corresponding to a sound speed of $c=0.2$ km/s,
  and adopt the scaling law Eq.~(\ref{equ:decay}) for decaying turbulence,
  we can express Jeans' criterion, Eq.~(\ref{equ:Jeans}), as
  \begin{equation}
  l_{c}^2>\frac{\pi}{\rho_c G} \bigg[c^2+\frac{1}{3}u_{c0}^2\bigg(1+\frac{A}{n}\frac{u_{c0}}{l_{c}}t\bigg)^{-n}\bigg],
  \label{equ:Jeans2}
  \end{equation}
  in which $u_{c0}$ denotes core turbulent speed at the beginning of turbulence decay.
According to the turbulence spectra of Eq.~(\ref{equ:kolmogorov}), the core turbulent speed is related to the core size by
  $u_{c0}=u_{0}(l_{c}/l_0)^{1/3}$, where $u_0$ and $l_0$ denote initial turbulent speed and size of the cloud,
  then Jeans' criterion (\ref{equ:Jeans2}) can be further expressed as
  \begin{equation}
  l_{c}^2>\frac{\pi}{\rho_c G} \bigg[c^2+\frac{1}{3}u_{0}^2\bigg(\frac{l_c}{l_0}\bigg)^{2/3}
  \bigg(1+\frac{A}{n}\frac{u_{0}}{l_{c}^{2/3} l_0^{1/3}}t\bigg)^{-n}\bigg].
  \label{equ:Jeans3}
  \end{equation}
Equation~(\ref{equ:Jeans3}) shows that the resistance to gravitational collapse in a cloud core has two contributions:
  the thermal and turbulent parts, which are respectively the first and second terms on the right hand side (RHS) of Eq.~(\ref{equ:Jeans3}).
For initially gravitationally stable cloud cores, the decay of the turbulent motion will diminish the second term
  and could cause the criterion being satisfied at a later time,
  hence providing an additional approach for the cloud core to become gravitationally unstable.

Therefore the existence of turbulence decay transforms Jeans' criterion, Eq.~(\ref{equ:Jeans3}), into two criteria:
  For a cloud core to be unstable at the beginning of the decaying turbulence ($t=0$),
  the core diameter has to be greater than
  \begin{equation}
    l_{c0}=\sqrt{\frac{\pi}{\rho_c G} \bigg[c^2+\frac{1}{3}u_{0}^2\bigg(\frac{l_c}{l_0}\bigg)^{2/3}\bigg]}.
  \label{equ:Jeans ini}
  \end{equation}
  On the other hand, if there is long enough time ($t\rightarrow +\infty$) for the turbulent motion to decay,
  the core diameter has only to be greater than
  \begin{equation}
  l_{c0}'=\sqrt{\frac{\pi}{\rho_c G} c^2}
  \label{equ:Jeans min}
  \end{equation}
  for it to be gravitational unstable and to collapse at some later time.
These two criteria can be inferred as the critical diameter for core collapse with initial turbulence ($l_{c0}$)
  and the minimum diameter required for core collapse without turbulence ($l_{c0}'$), respectively.
Note that the two criteria (\ref{equ:Jeans ini}) and (\ref{equ:Jeans min}) do not depend on $n$, the index of the turbulence decay rate,
  being only affected by the sound speed $c$ and the initial turbulent speed $u_{c0}=u_{0}(l_{c}/l_0)^{1/3}$ in cloud cores.
Based on these two criteria, three types of cloud-core evolution exist.
1) Cloud cores with diameters  $l_c>l_{c0}$ are gravitationally unstable and will collapse to form star(s).
2) Cloud cores with diameters in the range $l_{c0}'<l_c<l_{c0}$ will not collapse initially,
  but can evolve to be gravitationally unstable after a period
  of turbulence decay (turbulent speed decreases) and eventually collapse to proceed in further star formation.
3) Very small cloud cores with diameters $l_c<l_{c0}'$ that are stable and do not collapse.


\section{Cloud core relaxation}

When considering the time needed for a turbulent core to decay to being gravitationally unstable,
  it is more informative to re-write the criterion into the following form:
  \begin{equation}
  t>\frac{nl_{\rm c}}{Au_0(l_{\rm c}/l_0)^{1/3}}
  \bigg[\bigg(\frac{\frac{1}{3}u_0^2(l_{\rm c}/l_0)^{2/3}}{l_{\rm c}^2\rho_{\rm c}G/\pi-c^2}\bigg)^{1/n}-1\bigg].
  \label{equ:Jeans4}
  \end{equation}

For typical cloud cores with $c=0.2$ km/s and $\rho_c= 5\times10^4~{\rm cm}^{-3}$,
  if the initial turbulent speed of the (giant) cloud of diameter $l_0=100$ pc is $u_0=30$ km/s as estimated in Section 2,
  the time needed for a core of length scale $l_c$ to evolve to be gravitationally unstable
  can be easily obtained from Eq.~(\ref{equ:Jeans4}) and
  is illustrated in Fig.~1 (solid line), in which the decay index is taken as $n=1.2$.
Figure~1 shows that for cloud cores with sizes $l_c>l_{c0}= 3.0$ pc,
  $t=0$ yr, which means that these cores directly collapse as a result of gravity;
  while for those with sizes $l_c<l_{c0}'= 0.1$ pc, $t=+\infty$ yr, which means they will never experience gravitational collapse.
In between the two, cloud cores of sizes $l_{c0}'<l_c<l_{c0}$ become
  gravitationally unstable after a period of turbulence decay and collapse to form stars.
The epoch after the formation of these cores but before the initiation of gravitational collapse can be defined as the
  \emph{relaxation} of cloud cores, during which the turbulent intensity within the cloud cores decreases.
As inferred from Fig. 1 (solid line), the relaxation time for the cloud cores with sizes between $0.1$ pc and $3$ pc
  is around $10^5$ to $10^6$ yr,
  which is comparable to the free fall time for star formation, $t_{\rm ff}=[3\pi/(32G\rho_c)]^{1/2}=4\times10^5$ yr \citep{mckee2007}.
In addition, cloud cores with smaller sizes need longer relaxation times to become gravitationally unstable.


As the total energy released in a supernova explosion may not be fully converted to the turbulent energy of a star forming cloud,
  the initial turbulent speed of the entire cloud could be less than $30$ km/s as in previous calculation.
Observations also suggest that the turbulent speeds in clouds of $\sim 100$ pc diameter are typically less than or around $10$ km/s
  \citep[e.g.,][]{larson1981,mckee2007}.
Assuming 10\% or 1\% of the supernova explosion energy is converted to the turbulent energy of the molecular cloud,
  the corresponding turbulent speed is $u_0\sim10$ km/s and $u_0\sim3$ km/s, respectively.
Using Eq.~(\ref{equ:Jeans4}),
  the relaxation properties for dense cores in these less turbulent molecular clouds are also shown in Fig.~1.
The comparison between different cloud turbulent conditions shows that $l_{c0}$,
  the minimum diameter of a cloud core that can directly evolve to be gravitational unstable without experiencing turbulence decay,
  becomes much smaller when the initial turbulent speed decreases;
  while for cloud cores smaller than $l_{c0}$, the relaxation takes a relatively shorter time (several $10^5$ yr)
  than in more turbulent clouds.
The plots in Fig. 1 clearly indicate that the decay of turbulence leads to the existence of a relaxation epoch for
  cloud cores with diamater $l_{c0}'<l<l_{c0}$ before they experience gravitational collapse.
Even when Jeans' criterion has been satisfied and gravitational collapse begins,
  the decay of turbulent motion will also continue and the turbulent speed decreases until another driving process,
  such as star winds from nearby, newly formed massive stars, happens.
It is also to be noted that turbulence can be enhanced as a result of the adiabatic heating in the compression of a cloud core
  \citep{robertson2012,murray2014}.
This process cloud be considered as a self-driven mechanism of turbulence in cores as well,
  which may delay their gravitational collapses as a consequence.

Supernova driven turbulence has been presumed in the above analyses,
  while other sources reviewed in Section 2
  will also generate fluid turbulent motions.
Although in smaller scales and not energetic enough to be the main energy source for star formations \citep{maclow2004},
  these mechanisms may serve as more frequent energy inputs in local star formations.
In this sense, the core relaxation discussed above may be interrupted by these local turbulence drives.
Also note that magnetic field is not included in the analyses; the existence of which may lead to different
  turbulent energy spectra and may slow the decay of turbulence \citep[e.g.,][]{biskamp1999,mckee2007}.
Consideration of magnetic effects in future works may quantitatively change the turbulence decay and core relaxation behaviors discussed here.

\section{Conclusion and discussions}

Based on the scaling laws of decaying turbulence, Jeans' criterion on the stability of cloud cores specifies two critical core sizes:
  $l_{c0}$, if turbulence exists in the core, and $l_{c0}'$ ($<l_{c0}$), when only the thermal effect is considered.
For cloud cores with large enough sizes, ($l_c>l_{c0}$), they can be gravitationally unstable once formed.
For smaller cores that do not satisfy Jeans' criterion at their formation but have sizes between the two criteria ($l_{c0}'<l_c<l_{c0}$),
  they can evolve to be gravitationally unstable through the relaxation of turbulent energy.
For cores with even smaller sizes ($l_c<l_{c0}'$),
  they can never become unstable to gravity even with an infinite long epoch of relaxation.
The process of turbulence decay before gravitational collapse is defined as the relaxation of cloud cores,
  which lasts for a period of $10^5$ to $10^6$ yr for typical conditions in star forming clouds.
The existence of core relaxation provides an additional approach for cloud cores to evolve to be gravitationally unstable
  thus collapse.

Typical values of cloud core turbulent speed, length scale, density and temperature,
  as well as supernova rate and the (giant) cloud diameter are used here for
  an intuitive picture of the core relaxation;
  these values could vary from one star-forming cloud core to another by as large as even one or two orders of magnitude.

It is also noted that self-similar scaling laws of decay \citep{krogstad2011} and the
  Kolmogorov spectra of incompressible turbulence \citep{kolmogorov1941} are adopted here,
  with the analytical results apply in the ``inertial range"
  where energy transfers from larger to smaller scales with negligible influences from driving or viscosity.
Furthermore, although the existence of shock waves and the magnetic field in realistic molecular clouds may alter the turbulence spectra,
 the energy decay rates of turbulence for compressible and incompressible clouds with or without magnetic fields are quite comparable
 as found in numerical simulations \citep[see the reviews and discussions in][]{maclow2004,mckee2007}.
Nevertheless, the effects of compressibility, magnetic field as well as the anisotropy of turbulence
 on its decay properties, and consequently on the cloud-core relaxation needs to be further investigated.

\acknowledgments
This work was supported by the Center for Combustion Energy at Tsinghua University and by the National Science Foundation of China grant 51206088.
YG acknowledges additional support from the Tsinghua-Santander Program for young faculty performing research abroad.
HX acknowledges the support from the Max Planck Society and the German Science Foundation (DFG) through the project A7 of the Collaborative Research Center (CRC) 973 ``AstroFIT''.

\clearpage




\begin{figure}
 \epsscale{1} \plotone{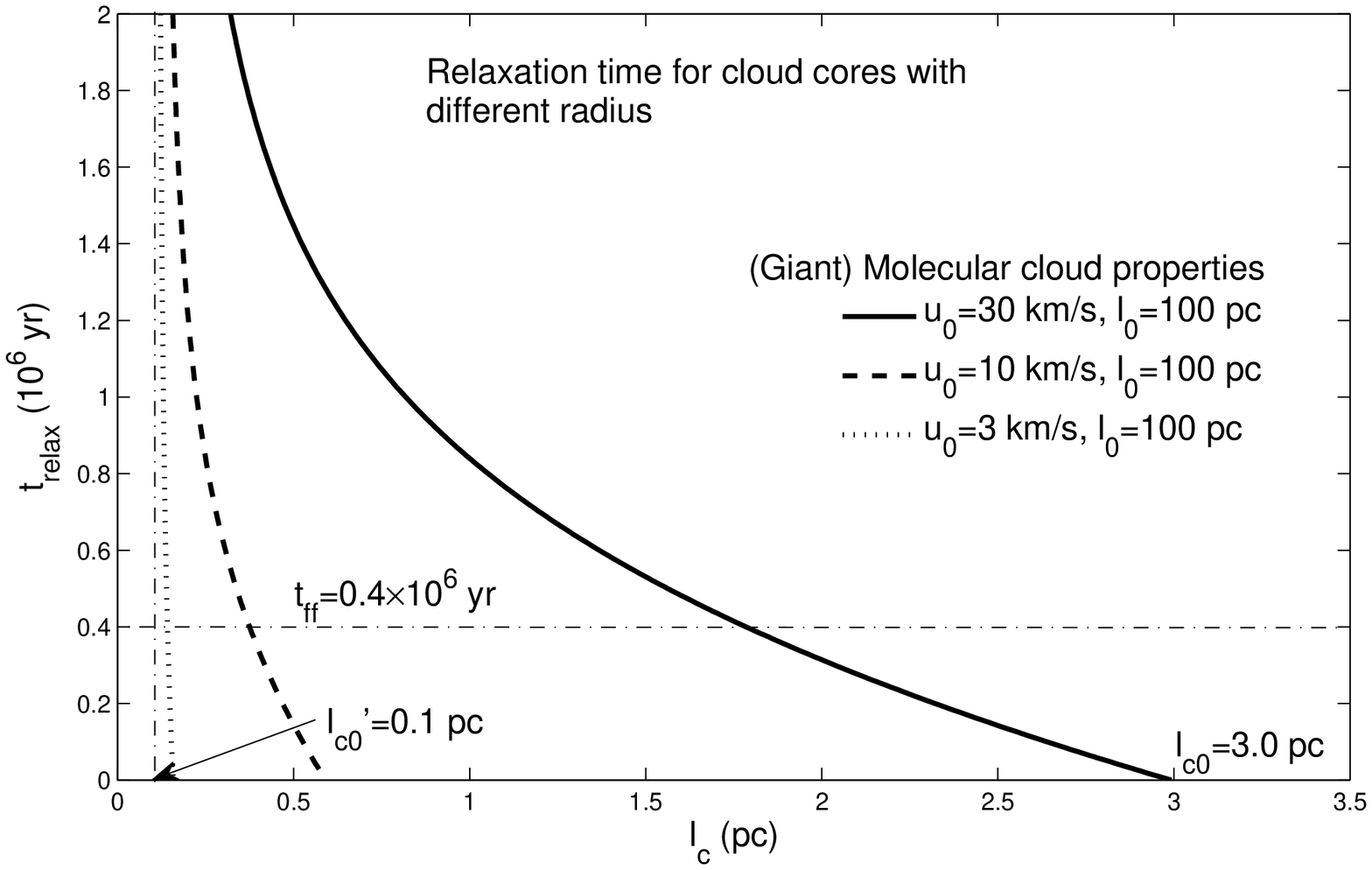}
\caption{Relaxation time needed for a cloud core embodied in a (giant) molecular cloud of diameter $l_0=100$ pc containing typical
  initial turbulent motions of $u_0=30$, 10 and 3 km/s to become gravitationally unstable.
  The abscissa is the scale of cloud core and the ordinate is the relaxation time needed before gravitational collapse.
For small cloud cores with $l_c<l_{c0}'$, the relaxation time is infinity and the core will always relax and can not become gravitationally unstable;
  for large cloud cores with $l_c>l_{c0}$, the relaxation time is zero and the core immediately goes to gravitational collapse once formed;
  for cloud core with diameter intermediate of the two, it needs a time $t_{\rm relax}$ for the turbulence to decay and
  eventually become gravitationally unstable.
The horizontal dash-dot line denotes the free fall time $t_{\rm ff}$ of the cloud core, which is comparable to the relaxation time.}
\end{figure}

\clearpage

\end{document}